\documentclass[prl,twocolumn,preprintnumbers,amsmath,amssymb,nofootinbib]{revtex4}
\usepackage{graphicx}% Include figure files
\usepackage{dcolumn}% Align table columns on decimal point
\usepackage{bm}% bold math

\usepackage{amsmath}
\usepackage{amsfonts}
\usepackage{amssymb}
\usepackage{feynmf}

\newcommand{\be}{\begin{equation}}
\newcommand{\ee}{\end{equation}}
\newcommand{\bear}{\begin{eqnarray}}
\newcommand{\eear}{\end{eqnarray}}
\newcommand{\beqstar}{\begin{eqnarray*}}
\newcommand{\eeqstar}{\end{eqnarray*}}

\newcommand{\mpl}{M_{\rm Pl}}

\newcommand{\Lammax}{ \Lambda_{\rm max} }
\newcommand{\rU}{ {\rm U} }
\newcommand{\cS}{ {\cal S} }
\newcommand{\Ssite}{ {\cS_{\rm site}} }
\newcommand{\Slink}{ \cS_{\rm link} }
\newcommand{\dx}{ {\rm d}^4 x }
\newcommand{\dvx}{ {\rm d}^5 x }
\newcommand{\dz}{ {\rm d} z }
\begin{document}
\begin{fmffile}{disc5pics}

\preprint{HUTP-03/A015}

\author{Nima Arkani-Hamed}
\email{arkani@schwinger.harvard.edu}
\author{Matthew D. Schwartz}
\email{matthew@schwinger.harvard.edu}
\title{Discrete Gravitational Dimensions}
% Force line breaks with \\

\affiliation{Jefferson Physical Laboratory, \\
Harvard University, Cambridge, MA 02138, USA}
% \altaffiliation[Also at ]{Physics Department, XYZ University.}
%Lines break automatically or can be forced with \\

%\date{\today}% It is always \today, today,
             %  but any date may be explicitly specified

\begin{abstract}
We study the physics of a single discrete gravitational
extra dimension using the effective field theory for massive
gravitons. We first consider a minimal discretization
with 4D gravitons on the sites and
nearest neighbor hopping terms.
At the linear level, 5D continuum
physics is recovered correctly, but at the non-linear level
the theory  becomes highly non-local in the
discrete dimension. There is a
peculiar UV/IR connection, where the scale of strong interactions
at high energies is related to the radius of the dimension. These new effects
formally vanish in the limit of zero lattice spacing, but do not
do so quickly enough to reproduce the continuum physics consistently
in an effective field theory up to the 5D Planck scale.
Nevertheless, this model does make sense as an effective theory up to
energies parametrically higher than the compactification scale. In order to
have a discrete theory that appears local in the continuum limit,
the lattice action must have interactions between distant sites. We
speculate on the relevance of these observations to the construction of
finite discrete theories of gravity in four dimensions.
\end{abstract}

\maketitle

In a recent paper
\cite{Arkani-Hamed:2002sp},
a technique was introduced for studying
gravitational
theories in discrete ``theory spaces'' 
\cite{Arkani-Hamed:2001ca, Hill:2000mu}.
These spaces are defined by sites, with
separate four-dimensional metrics, and their associated general coordinate
invariances, and ``link fields'' that map between sites.
In analogy with the
Callan-Coleman-Wess-Zumino formalism for gauge theories, this technique
allows us to
understand effective field theories with
multiple interacting spin two fields
in a transparent way. In the simplest case \cite{Arkani-Hamed:2002sp},
a single link is ``eaten'', leading to a single
massive graviton.
The physics of massive gravitons is qualitatively different than that
of massive gauge bosons, owing to the peculiar properties of the scalar
longitudinal component of the graviton. Nevertheless, there is a sensible
effective theory that makes sense to energies parametrically above the
graviton mass. In this letter, we study what happens when we
string together many sites and
links to generate what looks like a gravitational extra
dimension. Such models have been considered before
\cite{Damour:2002ws,
Sugamoto:2001uk,
Alishahiha:2001nb,
Bander:2001qk,
Jejjala:2002we,
Kan:2002rp},
but not at the level of a consistent effective field theory.
Given the peculiarities associated with massive gravity, we
should expect surprises, and we indeed encounter a number of them.

We will begin by considering the minimal discretizations,
with nearest neighbor interactions. The discretized dimension can be taken to
be either a circle or an interval:
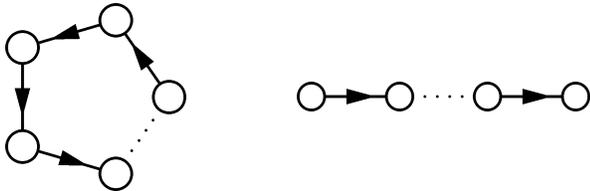
\begin{equation}
%
% 5 sites space
%
\parbox{40mm}{
\begin{fmfgraph*}(60,60)
\fmfsurroundn{v}{5}\fmfdotn{v}{5}
\fmfv{decoration.shape=circle,
decoration.filled=empty,decoration.size=0.2w}{v1}
\fmfv{decoration.shape=circle,decoration.filled=empty,
decoration.size=0.2w}{v2}
\fmfv{decoration.shape=circle,decoration.filled=empty,
decoration.size=0.2w}{v3}
\fmfv{decoration.shape=circle,decoration.filled=empty,
decoration.size=0.2w}{v4}
\fmfv{decoration.shape=circle,decoration.filled=empty,
decoration.size=0.2w}{v5}
\fmf{fermion}{v1,v2}
\fmf{fermion}{v2,v3}
\fmf{fermion}{v3,v4}
\fmf{fermion}{v4,v5}
\fmf{dots}{v5,v1}
\end{fmfgraph*} }
%
% 4 sites line
%
\parbox{40mm}{
\begin{fmfgraph*}(100,30)
\fmfleft{v1}
\fmfright{v4}
\fmf{fermion}{v1,v2}
\fmf{dots}{v2,v3}
\fmf{fermion}{v3,v4}
\fmfv{decoration.shape=circle,decoration.filled=empty,
decoration.size=0.1w}{v1}
\fmfv{decoration.shape=circle,decoration.filled=empty,
decoration.size=0.1w}{v2}
\fmfv{decoration.shape=circle,decoration.filled=empty,
decoration.size=0.1w}{v3}
\fmfv{decoration.shape=circle,decoration.filled=empty,
decoration.size=0.1w}{v4}
\end{fmfgraph*} }
\nonumber
\end{equation}
We will concentrate on the simpler case of an interval for now,
and return briefly to the circle later on.
Each of the $N$ sites is endowed with a four-dimensional
metric $g^j_{\mu \nu}(x)$.
The action
\begin{equation}
\cS = \Ssite + \Slink
\end{equation}
contains a part
\begin{equation}
\Ssite = \sum_j \int \dx M^2 \sqrt{g^j} R[g^j]
\end{equation}
which has $N$ copies of general coordinate invariance ($GC$). We can see this
by choosing a new dummy variable $x_j$ for each integral in the above sum.
The other part of $\cS$
involves interactions between neighboring sites, and will produce
mass terms of Fierz-Pauli form \cite{Fierz:1939ix}:
\begin{eqnarray}
\Slink^\rU = & \sum_j \int \dx \sqrt{g^j} M^2 m^2
(g^j_{\mu \nu} - g^{j+1}_{\mu \nu})
(g^j_{\alpha \beta} - g^{j+1}_{\alpha \beta}) &  \nonumber \\
&\times\quad  (g^{j \mu \nu} g^{j \alpha \beta} - g^{j \mu \alpha} g^{j \nu \beta})&
\label{linkU}
\end{eqnarray}
The U in $\Slink^\rU$
stands for Unitary gauge. Indeed, we can see that a gauge is chosen
because $\Slink^\rU$ breaks all but one copy of $GC$.

The action we have constructed is just
a naive discretization of compactified 5D Einstein gravity. To see this,
start with a 5D metric $G_{M N}(x,z)$, with $z$ the compact extra dimension.
Ignoring the radion and graviphoton degrees of freedom
(they do not affect the following discussion)
it is possible to choose a gauge where the metric takes the form
\begin{equation}
G_{M N} = \left( \begin{array}{cc} g_{\mu \nu} ( x , z ) & 0 \\
0 & 1 \\ \end{array} \right)
\end{equation}
In this gauge, the 5D action $\int \dvx M_{5D}^3 \sqrt{G} R^5[G]$ is
\begin{multline}
\cS = \int \dx \dz \sqrt{g} M_{5D}^3
\Big(  R_{4 D} [g] \\
+ \frac{1}{4} \partial_z g_{\mu \nu} ( g^{\mu \alpha} g^{\nu \beta}
- g^{\mu \nu} g^{\rho \beta} ) \partial_z g_{\alpha \beta} 
\Big) \label{Scont}
\end{multline}
Then, a naive discretization with lattice spacing $a$ instructs us to replace
\begin{equation}
\int \dz \to a \sum_j \, \, , \,  \partial_z g_{\mu \nu} \to \frac{1}{a}
(g^j_{\mu \nu} - g^{j+1}_{\mu \nu}) \label{contints}
\end{equation}
which directly produces our unitary gauge action \eqref{linkU}. And we can
read off that the radius $R$, lattice spacing $a$, and
effective 5D Planck scale $M_{5D}$ are given by:
\begin{equation}
R=N a, \quad
a=m^{-1}, \quad
M_{5D} = M^2 m \label{contvars}
\end{equation}

Now, as explained in \cite{Arkani-Hamed:2002sp},
we can restore the broken $GC$ symmetries in \eqref{linkU}
by adding link fields
$Y^\mu_{j}(x)$ between the sites.
It is useful to think of the link as a map
from a point on site
$j$ with coordinate $x^\mu_{j}$ to a point on site $j+1$ with coordinate
$Y^\mu_j(x_{j})$.
Under general coordinate transformations generated by $x_j \to f_j(x_j)$,
the link fields transform as:
\begin{equation}
Y_j \to f^{-1}_{j+1} \circ Y_j \circ f_{j}
\end{equation}
which allows us to compare fields on adjacent sites.

These link fields can be used to construct objects which transform as
tensors under $GC_{j}$ and are invariant under $GC_{j+1}$ out of objects
which are invariant under $GC_{j}$ and tensors under $GC_{j+1}$:
\begin{equation}
G^{j+1}_{\mu \nu}(x_j) \equiv
\partial_\mu Y_j^\alpha \partial_\nu Y_j^\beta g^{j+1}_{\alpha\beta}(Y_j(x_{j}))
\end{equation}
$G^{j+1}_{\mu \nu}(x_j)$ can be thought of as a pull-back of the metric
on site $j+1$ to site $j$ using the maps $Y_j$.
Thus, the link action becomes:
\begin{multline}
\Slink =
\sum_j \int \dx_j \sqrt{g^j} M^2 m^2
(g^j_{\mu \nu}(x_j) - G^{j+1}_{\mu \nu}(x_j)) \\
\times (g^j_{\alpha \beta}(x_j) - G^{j+1}_{\alpha \beta}(x_j))
(g^{j \mu \nu} g^{j \alpha \beta} - g^{j \mu \alpha} g^{j \nu \beta}) \label{linkG}
\end{multline}
We can now
replace the dummy variables $x_j$ by a common set of coordinates $x$.
By construction
this 4D action is explicitly invariant under $N$ copies of $GC$.

It is useful to expand the link fields about the identity as
\begin{equation}
Y_j^\mu(x) = x^\mu + \pi^\mu_j(x)
\end{equation}
The vector fields $\pi^\mu$ are the Goldstone bosons that are eaten in producing a
collection of massive spin two fields. Indeed, in unitary gauge we set 
$\pi_j=0$ ($Y_j^\mu(x)= x^\mu$)
and \eqref{linkG} reduces to \eqref{linkU}.

The spectum of this minimal theory space is that of standard latticizations:
there is a massless 4D graviton, and
tower of massive spin two fields, with a characteristic
lattice spectrum $m_n = m \sin(n/N)$. For $n\ll N$ the spectrum approaches
a KK tower of compactified 5D theory.
At the linear level, the exchange of these modes
generates the correct 5D graviton propagator up to small corrections, and
so, for example, 5D Newtonian gravity is reproduced.

However, at the
non-linear level peculiar new features are revealed, which can be traced to
the interactions of the longitudinal modes of the massive gravitons,
as in
\cite{Arkani-Hamed:2002sp}.
Expanding the metrics about flat
space as $g^j_{\mu \nu} = \eta_{\mu \nu} + h^j_{\mu \nu}$, the hopping terms
involve
\begin{equation}
g^{j}_{\mu \nu} - G^{j+1}_{\mu \nu} = h^{j}_{\mu \nu} - h^{j+1}_{\mu \nu} +
\pi^{j}_{\mu,\nu} + \pi^{j}_{\nu,\mu} +
\pi^{j \alpha}_{,\mu} \pi^j_{\alpha,\nu} + \cdots \label{gminusG}
\end{equation}
where the $\cdots$ refer to terms involving more powers of the $h$.
As in the study of a single massive graviton, it is useful to decompose the
$\pi^\mu_j$ as
\begin{equation}
\pi^j_\mu = A^j_\mu + \partial_\mu \phi^j
\end{equation}
The dynamics of the $\phi^j$ is that of the scalar longitudinal components
of the
gravitons, and they produce the amplitudes that grow most dangerously with
energy.

Inserting \eqref{gminusG} into \eqref{linkG} and going
to momentum space
displays the kinetic terms and interactions for all the
Goldstone modes. Schematically:
\begin{multline}
\cS  = 
\int \dx N M^2 h_n\partial^2 h_n
+ N M^2 m^2 \Big(\frac{n^2}{N^2} h_n^2  \\
+ \frac{n}{N} h_n \partial^2 \phi_n
+ (\partial^2\phi_n)(\partial^2 \phi_m)(\partial^2\phi_{n+m})
+\cdots \Big) \label{snN}
\end{multline}
where $h_0$ is the massless graviton and $h_n$ and $\phi_n$
are the graviton and scalar Goldstone
at the $n^{th}$ mass level.
Now, $\phi_n$ picks up a kinetic term
of the form $M^2 m^4 n^2 N^{-1} \phi_n \partial^2 \phi_n$
from mixing with $h_n$  and the
strongest interactions come from the $\partial^6 \phi^3$ vertex
for the lowest modes,
just as in \cite{Arkani-Hamed:2002sp}.
For instance the amplitude for
$\phi_1$-$\phi_1$ scattering goes as:
\begin{equation}
{\cal A} \quad=\quad
%
% o o -> o -> o o
%
\parbox{30mm}{
\begin{fmfgraph*}(85,40)
\fmfleft{p1,p2}
\fmfright{q1,q2}
\fmf{dashes}{p1,pl}
\fmf{dashes}{p2,pl}
\fmf{dashes}{pl,pr}
\fmf{dashes}{pr,q1}
\fmf{dashes}{pr,q2}
\fmfv{l=$\phi_1$,l.a=120,l.d=.03w}{p1}
\fmfv{l=$\phi_1$,l.a=-120,l.d=.03w}{p2}
\fmfv{l=$\phi_1$,l.a=60,l.d=.03w}{q1}
\fmfv{l=$\phi_1$,l.a=-60,l.d=.03w}{q2}
\end{fmfgraph*} } \quad  \sim \quad
\frac{E^{10}}{\Lambda^{10}} \label{scalscat}
\end{equation}
Where, expressed in terms of the low
energy 4D Planck scale $\mpl = \sqrt{N} M$ and the
mass of the first KK mode $m_1 =m/N$:
\begin{equation}
\Lambda = (N m_1^4 \mpl)^{1/5} \label{LminN}
\end{equation}
This is higher by a factor of $N^{1/5}$ than the
scale that the theory of a single graviton of mass $m_g=m_1$
breaks down. 
%Note that if we scattered the $n^{th}$ mode
%instead of the first, the strong coupling scale would be higher
%(\eqref{LminN} with $m_n$ replacing $m_1$).

In terms of 5D variables:
\begin{equation}
\Lambda = \left(\frac{M_{5D}^{3}}{R^{5} a^{2}}\right)^{1/10}\label{Lmin}
\end{equation}
Note that bizarrely, the UV scale at which the theory becomes
strongly coupled depends on an IR scale, the
{\it size} of the extra dimension!

Naturally, as we
decrease the lattice spacing, $\Lambda$ increases. However, for a
consistent effective theory, we should require that all the states in the
theory be lighter than the UV cutoff $\Lambda$. In particular, the
heaviest KK mode, of mass $\sim a^{-1}$, should be lighter than $\Lambda$.
This means that the lattice spacing cannot be decreased beyond a certain
point, and the highest UV cutoff the theory can possibly have is
\begin{equation}
\Lammax \sim a^{-1}_{\rm min} \sim \left(\frac{M_{5D}^3}{R^5}\right)^{1/8}
\end{equation}
Again, this strikingly exhibits a UV/IR connection: the highest possible
UV cutoff $\Lammax$ decreases as the size $R$ of the dimension is
increased in such a way that
$\Lammax^8 R^5 = M_{5D}^3$ stays fixed. Note that for any radius larger
than the 5D Planck length, $\Lammax$ is always smaller
than $M_{5D}$.
In other words, the minimal lattice cannot look like 5D gravity
at the non-linear level all the way up to the 5D Planck scale.

It is useful to understand what is going on directly in unitary gauge,
where the amplitude \eqref{scalscat} is that of scattering
scalar longitudinal ($sL$) polarizations of the lightest massive
gravitons.
The amplitude \eqref{scalscat} can be written in the instructive form:
\begin{equation}
{\cal A} \sim \frac{E^{10}}{\Lambda^{10}} \sim \frac{E^{10} R^5 a^2}{M_{5D}^3}
\sim \frac{E^{10}}{\mpl^2 (1/R)^8} \times \frac{a^2}{R^2} \label{AofR}
\end{equation}
In the case of a single
massive graviton of mass $m_g$ and Planck scale $\mpl$, there are two
contributions to the amplitude for graviton scattering, one from
graviton exchange and one from the direct 4-point graviton vertex
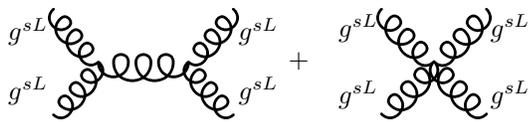
\begin{equation}
%
% o o -> o -> o o
%
\parbox{30mm}{
\begin{fmfgraph*}(85,40)
\fmfleft{p1,p2}
\fmfright{q1,q2}
\fmf{gluon}{p1,pl}
\fmf{gluon}{p2,pl}
\fmf{gluon}{pl,pr}
\fmf{gluon}{pr,q1}
\fmf{gluon}{pr,q2}
\fmfv{l=$g^{sL}$,l.a=120,l.d=.05w}{p1}
\fmfv{l=$g^{sL}$,l.a=-120,l.d=.05w}{p2}
\fmfv{l=$g^{sL}$,l.a=60,l.d=.05w}{q1}
\fmfv{l=$g^{sL}$,l.a=-60,l.d=.05w}{q2}
\end{fmfgraph*} } \quad +\quad\quad
%
% o o ->  o o
%
\parbox{25mm}{
\begin{fmfgraph*}(50,40)
\fmfleft{p1,p2}
\fmfright{q1,q2}
\fmf{gluon}{p1,p}
\fmf{gluon}{p2,p}
\fmf{gluon}{p,q1}
\fmf{gluon}{p,q2}
\fmfv{l=$g^{sL}$,l.a=120,l.d=.05w}{p1}
\fmfv{l=$g^{sL}$,l.a=-120,l.d=.05w}{p2}
\fmfv{l=$g^{sL}$,l.a=60,l.d=.05w}{q1}
\fmfv{l=$g^{sL}$,l.a=-60,l.d=.05w}{q2}
\end{fmfgraph*} }
\nonumber
\end{equation}
There is no cancellation between these two contributions, so the amplitude grows as
$E^{10}/(m_g^8 M_{Pl}^2)$. Our scattering amplitude has exactly the same form,
with $m_g \to 1/R$, however there is a suppression factor of
$a^2/R^2$. Evidently, in the continuum theory, there is an {\it exact}
cancellation between the two contributions, ensured by the 5D
gravitational Ward identities. However, in the discretized
theory, the spectrum and interactions are modified by $\sim (a/R)$ effects,
and so the cancellation is imperfect, reflecting the breaking of
the 5D general coordinate invariance by our discretization.

Needless to say, this behavior is dramatically different than for gauge theories.
The same theory space for a non-Abelian gauge theory would become strongly
coupled at an energy scale which is {\it always} higher than than the mass of all the
modes. So the discretized theory can be made to look identical to a higher
dimensional gauge theory all the way up to scale where the 5D (non-renormalizable)
gauge theory would naturally break down.

Nevertheless, the minimal gravitational model
does define a sensible effective field theory of gravity valid to energies
parametrically above the compactification radius $1/R$.
The strong interactions formally vanish in the limit of zero lattice
spacing, they just do not vanish quickly enough to reproduce
5D gravity all the way up to $M_{5D}$ in a
consistent effective field theory.

Let us now study this model in position space, where
the physics is more transparent. Working directly in continuum language,
the action $\cS$ becomes:
\begin{multline}
\cS = \int \dx \dz 
M_{5D}^3 \Big( 
(\partial h)^2 + (\partial_z h)^2 \\
+ \frac{1}{a} h \partial^2 \partial_z \phi + \frac{1}{a^2}
(\partial^2 \phi)^3 + 
\cdots \Big)
\end{multline}
%This can be derived from plugging the continuum variables \eqref{contints}
%and \eqref{contvars} into \eqref{snN} or simply by substituting
%$a \partial_z h \to a \partial_z h + \partial^2 \phi + (\partial^2\phi)^2$
%into the linearized 5D Lagrangian.
We have done an integration by parts to bring the bilinear mixing between
$h$ and $\phi$ to the above form. 
Note also that because the extra dimension is an
interval, the boundary conditions at the ends are that $\partial_z h$ and
$\pi$ vanish, so that there are no Goldstone zero modes (corresponding to
the fact that they are all eaten).

We can remove the kinetic mixing between $h$ and $\phi$ by defining
\begin{equation}
h_{\mu \nu} = \hat{h}_{\mu \nu} - \eta_{\mu \nu} \psi, \, \, \mbox{where}
\, \, \psi = \frac{1}{a} \partial_z \phi
\end{equation}
which generates a kinetic term for $\phi$. We add gauge-fixing terms
directly for $\hat{h}$, the precise form of which will not be important
here. The $\phi$ action then becomes:
\begin{equation}
\cS = \int \dx \dz M_{5D}^3
\left((\partial \psi)^2 + (\partial_z \psi)^2 
+ \frac{1}{a^2}(\partial^2 \phi)^3 + \cdots \right) \nonumber
\end{equation}
%where we have omitted higher order terms in $\phi$ and terms involving
%$\hat{h}$.

Already we see that this action is strange. The combination $\psi =
\frac{1}{a} \partial_z \phi$ has a normal kinetic term. $\psi$ also
couples directly to the trace of the energy momentum tensor in the
way required to produce the correct tensor structure for the propagator
in 5D.
So, at the linear level, $\psi$ is the physical propagating field.
Observe, however, that the self-interactions involve $\phi$
and are therefore highly non-local with respect to
$\psi$. To see this, we can formally
write $\phi = \frac{a}{\partial_z} \psi$; with the boundary conditions that
$\phi$ vanishes at the ends of the interval, $\partial_z$ is invertible and
this can be done unambiguously. The interaction Lagrangian for $\psi$ is
then
\begin{equation}
\int \dx \dz M_{5D}^3 a (\frac{\partial^2}{\partial_z} \psi)^3 + \cdots \label{psi3}
\end{equation}
which is manifestly non-local in the $z$ direction.

Note that the interaction \eqref{psi3} formally
vanishes in the zero lattice spacing limit, just as $\Lambda\to\infty$ as
$a\to0$ in \eqref{Lmin}.
However for any finite lattice spacing
$a$, at large enough distances this term can become important.
On a finite interval of size $R$, the largest wavelength modes of size $\sim
R$ will suffer the strongest interactions. These modes $\sim \psi_R$ correspond to the
longitudinal polarizations of the lowest KK modes, and are described by the effective
4D action:
\begin{equation}
\int \dx \Big( M_{5D}^3 R  (\partial \psi_R)^2 
+ M_{5D}^3 a R^4 (\partial^2 \psi_R)^3 + \cdots \Big)
\end{equation}
We can see immediately that the amplitude for
$\psi_R \psi_R \to \psi_R \to \psi_R \psi_R$ is the same as \eqref{scalscat}.
It is precisely the non-local interactions of $\psi$ which lead
to the strong amplitudes which force $\Lambda \ll M_{5D}$.

We have seen that, while our starting point {\it appears} extremely
local, with only nearest neighbor hopping terms for the gravitons on the
sites, it actually induces highly non-local interactions in the discretized
dimension.
Why did this happen? Presumably, these effects are related to the fact that
we have broken the full 5D diffeomorphism invariance of the theory by our
discretization. Usually, however, when gauge symmetries are explicitly
broken, a theory gains new degrees of freedom (which correspond to pure
gauge modes in the gauge invariant theory). Here (ignoring the irrelevant
radion), the number of physical degrees of freedom beneath $a^{-1}$ match
exactly between the discrete and continuum theories. Indeed, despite the
absence of the full 5D diffeomorphism invariance and Lorentz invariance,
the theory still has an exactly ``massless'' graviton even in the 5D sense,
by which we mean gapless excitations with $\omega \to 0$ as $k \to 0$.
However, the {\it
interactions} are qualitatively different in the two theories.

So far we have discussed compactification on an interval. When we compactify
on a circle, we have to contend with the Goldstone zero mode, which is
no longer killed by the boundary conditions.
The zero mode has $\psi_0 =a^{-1}\partial_z \phi_0=0$ (equivalently,
the $n=0$ mode in \eqref{snN})
and therefore has no kinetic term at all. However, it
does appear in interactions! This means that it is inconsistent not to
include a kinetic term for $\phi_0$ in the theory, which becomes a mass
term for the graviphoton $\pi_0^\mu$ (more specifically, the generally
covariant plaquette
operator
\cite{Arkani-Hamed:2002sp} made from the ``Wilson line'' $Y_1 \circ \cdots
\circ Y_N$ is generated, which contains the mass term for
$\pi^0_\mu$ as well as
various non-linear interactions).
It is not surprising that such a term should be generated.
In the continuum, the massless graviphoton is associated with a $U(1)$ gauge
symmetry inherited from the 5D reparameterization invariance of the circle.
This invariance is clearly broken by the discretization. We can see this
concretely, because in the continuum $\phi_0$ is a pure gauge mode with no
dynamics at all, while in our discretization it does appear in
interactions. Therefore, the $U(1)$ gauge invariance is explicitly broken
and there is no symmetry to prevent $\pi^0_\mu$ from picking up a mass.

A similar analysis can be done for the case where we start with 3D sites
and go up to a 4D theory. Amusingly, in this case,
the 3D gravity on the sites does
not have any propagating degrees of freedom --  all of the local
degrees of freedom in the 4D theory come from the links. Once again, the
$\phi$ fields only acquire a kinetic term by mixing with the site metrics,
and the same sorts of non-local interactions arise. The
maximum cutoff for a consistent effective theory in this case is
\begin{equation}
\Lammax \sim \left(\frac{\mpl^2}{R^5}\right)^{1/7}
\end{equation}
Of course, the discretization of one of three spatial dimensions in our
universe leads to the breaking of rotational invariance and there are
already very severe limits on such effects. But even ignoring this
the above cutoff is still incredibly low; taking $R \sim 10^{28}$ cm to be
about the size of the universe today, $\Lammax^{-1} \sim 10^{12}$ cm!
Clearly such discretizations are not sensible to describe our 4D world.

Are there discretizations that avoid inducing non-local interactions and
correctly reproduce
local continuum physics? Simply adding
next-to-nearest neighbor interactions, or anything similar which
is essentially local on
the lattice, cannot possibly work.
We can get a hint of
what is needed by looking at a discretization that is {\it guaranteed}
to work, at least at tree-level.
Take a continuum theory on an interval and simply {\it truncate} the KK tower,
keeping only $N$ of the modes. Of course, we do not need an infinite
number of states \cite{Duff:ea} for an effective theory.
By KK momentum conservation, the tree-level
scattering amplitudes for the lowest KK
modes
cannot involve the truncated modes,
and therefore their scattering
amplitudes coincide with the (healthy) ones of the continuum theory.
The strongest tree level amplitudes come from the scattering of the
modes between $\sim N/2$ and $N$ near the top of the tower, of mass
$\sim N/R$.
As is discussed in detail in \cite{Schwartz:2003vj}
the strong coupling scale in this case is  determined by the $\Lambda_3$
scale 
%\cite{Arkani-Hamed:2002sp}
associated with this mass:
\begin{equation}
\Lambda \sim \left((\frac{N}{R})^2 \mpl \right)^{1/3}
\sim \left(\frac{M_{5D}^3 R}{a^4} \right)^{1/6}
\end{equation}
So, for any lattice spacing $a$, the
model makes sense as an effective theory up to energies
parametrically higher than the top of the KK tower. Therefore in contrast
with our minimal discretization, we can take a
limit where the new strong amplitudes induced by the discretization are no
more important than those associated with the 5D Planck scale $M_{5D}$. At
least at tree-level, this model can make sense all
the way up to $M_{5D}$, reproducing local 5D physics.

Starting
with these $N$ modes in momentum space, we can go back to an $N$-site
theory in position space by a discrete Fourier transform.
The sharp momentum truncation
induces interactions between sites in position space that are
not strictly local:
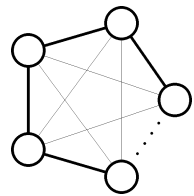
\begin{equation}
%
% 5 sites space
%
\parbox{20mm}{
\begin{fmfgraph*}(60,60)
\fmfsurroundn{v}{5}\fmfdotn{v}{5}
\fmfv{decoration.shape=circle,
decoration.filled=empty,decoration.size=0.2w}{v1}
\fmfv{decoration.shape=circle,decoration.filled=empty,
decoration.size=0.2w}{v2}
\fmfv{decoration.shape=circle,decoration.filled=empty,
decoration.size=0.2w}{v3}
\fmfv{decoration.shape=circle,decoration.filled=empty,
decoration.size=0.2w}{v4}
\fmfv{decoration.shape=circle,decoration.filled=empty,
decoration.size=0.2w}{v5}
\fmf{plain}{v1,v2}
\fmf{plain}{v2,v3}
\fmf{plain}{v3,v4}
\fmf{plain}{v4,v5}
\fmf{dots}{v5,v1}
\fmf{plain,width=0.15}{v1,v3}
\fmf{plain,width=0.15}{v2,v4}
\fmf{plain,width=0.15}{v3,v5}
\fmf{plain,width=0.05}{v2,v5}
\fmf{plain,width=0.05}{v1,v4}
\end{fmfgraph*} }
\nonumber
\end{equation}
The Fourier transform of a step function
has a rapid oscillatory behavior, but dies off as a {\it power}
with distance in position space \cite{Schwartz:2003vj}, 
rather than exponentially as in genuinely
local theories.

%\section{Summary and outlook}

\end{fmffile}

The non-local interactions we found in the minimal discretization
have the same origin as the strong-coupling effects for
a single massive graviton discussed in
\cite{Arkani-Hamed:2002sp}. These strong coupling effects were softened
in curved backgrounds (such as AdS), so it would
interesting to look for similar improvements in
the context of discrete dimensions, where the sites are taken to have AdS
geometries. Another straightforward exercise is to
discretize warped geometries (such as AdS$_5$). Since quantum gravity in
AdS$_5$ is dual to a 4D field theory, it would be interesting to see whether
a naive discretization can do a better job in describing the continuum
physics in this case, at least at scales larger than the AdS$_5$ length.
%\cite{Randall:2002tg}.

We have seen that a naive local discretization of gravity induces
non-local interactions at long distances, in sharp contrast to gauge
theories, where the minimal discretization is perfectly well-behaved.
This is apparent already at tree-level for a single discrete dimension,
and makes it impossible to reproduce the correct continuum physics within
a consistent effective field theory.  We have also seen that a less local
discretization, comprising very specific interactions between distant
sites in position space, does better than the minimal discretization, and
can successfully reproduce local physics at low energies at least at
tree-level. This discretization follows from a simple truncation of the
continuum Kaluza-Klein tower.  In gauge theory, such a truncation is
artificial and the corresponding non-local interactions in position space
give rise to various pathologies; it is interesting that in the gravity
case this ``artificial" discretization is superior to the minimal one.

Ultimately, it is of interest to consider full space or space-time lattices
that may provide a sensible definition of quantum gravity in four
dimensions.  However, despite the many hints of discrete structures
underlying quantum gravity, lattice approaches have all suffered from the
inability to reproduce the correct continuum physics at low energies. Our
simple examples suggest that a successful discretization of gravity cannot
look local, but should involve a special set of non-local interactions
between distant sites. It is tempting to speculate that this is a
reflection of a UV/IR connection in quantum gravity, and that the
overly-local nature of most approaches to lattice gravity is responsible
for their failures at low energy. Of course, further progress requires
finding a principle that dictates the structure of the required non-local
interactions.

We would like to thank 
Hsin-Chia Cheng,
Paolo Creminelli, 
and especially
Howard Georgi
for many illuminating discussions. We also thank Michael Graesser and
Christophe Grojean for pointing out important typos in some equations of
the first version of this note. 
The work of N. A-.H. is partially
supported by the David and Lucille Packard foundation, and the Alfred
P. Sloan foundation.

\end{document}